# Improving the glial differentiation of human Schwann-like adipose-derived stem cells with graphene oxide substrates

*Andrea Francesco Verre,*[1] *Alessandro Faroni,*[2] *Maria Iliut,*[1] *Claudio Silva,*[1,3] *Cristopher Muryn,*[4] *Adam J Reid* [2,5] *and Aravind Vijayaraghavan*[1,*]

[1] School of Materials and National Graphene Institute, University of Manchester, Manchester M13 9PL, UK

[2] Blond McIndoe Laboratories, Division of Cell Matrix Biology and Regenerative Medicine, School of Biological Sciences, Faculty of Biology Medicine and Health, University of Manchester, Manchester Academic Health Science Centre, Manchester M13 9PL

[3] Department of Fundamental Chemistry, Institute of Chemistry, University of São Paulo, São Paulo, Brazil.

[4] School of Chemistry, University of Manchester, Manchester M13 9PL, UK

[5] Department of Plastic Surgery & Burns, University Hospitals of South Manchester, Manchester Academic Health Science Centre, Manchester

* Corresponding author: aravind@manchester.ac.uk

**Abstract**

There is urgent clinical need to improve the clinical outcome of peripheral nerve injury. Many efforts are directed towards the fabrication of bioengineered conduits, which could deliver stem cells to the site of injury to promote and guide peripheral nerve regeneration. The aim of this study is to assess if graphene and related nanomaterials can be useful in the fabrication of such conduits. A comparison is made between GO and reduced GO substrates. Our results show that the graphene substrates are highly biocompatible, and the reduced GO substrates are more effective in increasing the gene expression of the biomolecules involved in the regeneration process compared to the other substrates studied.

## 1. Introduction

Schwann cells (SC) are key cellular elements in assisting the regeneration of peripheral nerve after injury. SC switch from a myelinating to repair phenotype which results in increased expression of extracellular matrix (ECM) proteins, neurotrophins and growth factors; furthermore, SC undergo profound morphological changes which result in upregulation of filament cytoskeletal proteins such as nestin and actin [1,2]. Despite a clear need for novel therapies, use of SC as a clinical intervention for peripheral nerve injury (PNI) is problematic due to the necessity of harvesting a functional nerve and the limited expansion capacity of SC. As a clinically viable alternative, mesenchymal stem cells and adipose-derived mesenchymal stem cells (ASCs) have been differentiated *in vitro* towards a Schwann-like cells phenotype [3,4]. These differentiated adipose stem cells (dASCs) express glial markers such as glial fibrillary acidic protein (GFAP), S100 and p75 [4]; express myelin protein [5] and myelin structures when in co-culture with neurons [6,7]; and when implanted in bioengineered conduits to repair murine peripheral nerve gap *in vivo*, dASC have demonstrated promotion of nerve regeneration, reduction of muscle atrophy, increased nerve conduction velocity and higher rates of myelination [8-11].

Graphene and related nanomaterials can play an important role in the fabrication of bioengineered nerve conduits for the treatment of peripheral nerve injuries. Although the biocompatibility of these materials for *in vivo* studies depends on many variables such as the thickness, the lateral size of the flakes, the level of hydrophilicity and the extent of functionalization [12], it can be stated that when these materials were used as coatings on surfaces to support stem cell growth, the extent of cytotoxicity was limited and enhanced stem cell differentiation was reported [13-17]. Graphene and related nanomaterials were found to be effective in positively modulating axonal outgrowth and nerve regeneration *in vitro* [18-21]. Thus far, researchers have been exploring the effect of graphene and related nanomaterials on the neurite outgrowth, but there have not been studies regarding the effect of these materials in supporting the growth of dASCs. The aim of this study is then the biological characterization of graphene oxide (GO) and reduced GO (rGO) coated coverslips and verify if these materials are able to sustain the differentiated dASCs phenotype, which is rapidly lost following withdrawal of growth factors.

## 2. Materials and Methods

### 2.1 Graphene based materials synthesis, substrates preparation and characterization

Graphite oxide was synthesized by a modified Hummer`s method [22,23] and exfoliated down to constituent monolayers to yield GO. Glass coverslips were washed in a sonication bath with Decon® 90 for 15 mins followed by deionised (DI) water for another 15 mins and finally with isopropanol for 15 mins. After the coverslips were dried, they were treated for 5 mins in oxygen plasma to increase the hydrophilicity of the coverslips. GO dispersions were then spin-coated on the glass coverslips at the concentration of 2 mg/ml at 2500 rpm, 250 rpm/sec acceleration for 2 mins. To obtain rGO coverslips, GO coverslips were kept for three days at 180 °C in vacuum. Atomic force microscopy (AFM) measurements were carried using a Bruker FastScan microscope in tapping mode. Raman spectroscopy was performed using a Renishaw inVia Raman microscope with 532 nm laser excitation. X-ray photoelectron spectroscopy (XPS) spectra of drop-casted substrates were recorded with a SPECS NAP-XPS system employing a monochromatic Al Kα source (1486.6 eV). Before the biological experiments, the coverslips were sterilised by immersion in pure ethanol then washed in deionised water and then left to dry under the hood.

### 2.2 Human Adipose stem cells harvesting and differentiation

ASCs were isolated according to a previously reported protocol [4]. Human abdominal, subcutaneous adipose tissue was harvested from three female surgical patients undergoing reconstructive surgery at the University Hospital of South Manchester, UK. All patients were fully consented and procedures approved by the National Research Ethics Committee, UK (NRES 13/SC/0499). Adipose tissue biopsies were minced by a razor blade and dissociated by an enzymatic treatment of 0.2 % (w/v) of collagenase I (Life Technologies, Paisley, UK) for 60 minutes at 37 °C under constant agitation. The digested tissue was then filtered through a vacuum-assisted 100 µm nylon mesh (Merck Millipore, Watford, UK). An equal volume of stem cell growth medium containing a-minimum essential Eagle's medium (aMEM) (Sigma-Aldrich, Poole, UK), 10% (v/v) foetal bovine serum (FBS) (LabTech, Uckfield, UK), 2 mM L-glutamine (GE Healthcare UK, Little Chalfont, UK), and 1% v/v penicillin–streptomycin was added. The solution were centrifuged at 300 g for 10 minutes and the resulting pellet was suspended in 1 mL of Red Blood Cell Lysis Buffer (Sigma-Aldrich) for 1 min, and 20 mL of aMEM was added to arrest lysis. The mixture was centrifuged at 300 g for 10 min, and the resulting pellet was resuspended in aMEM and plated in T75 flasks for cell culture. Cells were routinely characterised for the expression of stem cell surface markers as per [24]

The differentiation of ASC towards dASCs was performed following a previously reported protocol [4]. Briefly, ASCs at the passage 1-2 at 30% of confluence were treated with 1 mM β-mercaptoethanol (sigma-Aldrich) for 24 hours, then with 35 ng/mL of all-*trans*-retinoic acid for 72 hours. After this initial treatment, ASCs were treated by 5 ng/mL of platelet-derived growth factor (Peprotech EC, London,UK), 10 ng/mL basic fibroblast growth factor (Peprotech EC), 14 μM of forskolin (Sigma-Aldrich) and 192 ng/mL glial growth factor (GGF-2) (Acorda Therapeutics, Ardsley, NY, USA). ASCs were kept under these conditions for 2 weeks, replacing media every 72 hours and passaging when confluence was reached.

## 2.3 Cell Proliferation and Live/Dead assays

To assess cell proliferation rate by the (3-(4,5-dimethylthiazol-2-yl)-5-(3-carboxymethoxyphenyl)-2-(4-sulfophenyl)-2H-tetrazolium) (MTS) assay , dASCs cells were plated at the concentration of 5,000 cells per coverslip in triplicate on the sterilised graphene-coated coverslips. The coverslips were put in ultra-low adherence 24-well cell plates to avoid attachment of the cells to the tissue culture plastic underneath the coverslips. At days 1, 4 and 7 after seeding the cell medium was aspirated and cells were washed in phosphate buffered saline (PBS). After the washing step, the cells were incubated in 20% (v/v) CellTiter 96 Aqueous One Solution Cell Proliferation Assay (Promega, Southampton, UK), diluted in phenol-free DMEM (Sigma-Aldrich) for 90 mins in the dark at 37 °C. After the incubation, the absorbance at 490 nm was recorded using an Asys UVM-340 microplate reader/spectrophotometer (Biochrom, Cambridge, UK).  For the viability assays, ASC cells were plated at the concentration of 25,000 cells per coverslips in triplicate in a cell medium containing α-MEM and 1% (v/v) P/S without growth factors. After 48 hours, the medium was aspirated and the cells were washed in PBS. Calcein-AM fluorescein and ethidium homodimer-1 (Eth-D1) purchased from LIVE/DEAD® Viability/Citoxicity Kit (Molecular Probes, Invitrogen, UK) were added at the concentration of 0.5 and 2 μg/ml respectively in PBS and left to react for 15 minutes at 37º C. After this step, images were taken using a fluorescence inverted microscope (Olympus IX51, Japan) under 4X magnification. Data from MTS experiment were expressed were expressed as absorbance at 490 nm ± SE of the mean (n = 3), while data from viability assay were expressed as percentage of live cells measured by dividing the average green stained area by the average of the whole (Live + Dead) stained area, measured with Image J software (version 1.48) multiplied by 100.

**2.4     Quantitative real-time polymerase chain reaction (qRT-PCR)**

For gene expression studies, cells were seeded as above at a concentration of 50,000 cells per coverslips in triplicate and the RNA was extracted after 48 hours of cellular growth on the different coverslips. RNA was extracted using the RNeasy Plus Mini Kit (Qiagen) following the instruction of the manufacturer. The concentration of the RNA was quantified at the NanoDrop ND-100 (Thermo Fisher Scientific, Waltham, MA, USA) spectrophotometer. 1 µg of each sample were reverse transcribed using the RT2 First Strand Kit (Qiagen) following the instruction of the manufacturer. DNA elimination steps were included in both RNA extraction and cDNA synthesis to prevent downstream genomic DNA amplification. qRT-PCR was performed with RT2 SYBR Green qPCR Mastermix (Qiagen) and a Corbett Rotor Gene 6000 Qiagen), by the use of the following protocol: hot start for 10 min at 95 °C, followed by 40 cycles of 15 s at 95 °C, annealing for 30 s at 55 °C, and extension for 30 s at 72 °C. To verify the specificity of the reactions, a melting curve was obtained with the following protocol: 95 °C for 1 min, 65 °C for 2 min, and a gradual temperature increase from 65 °C to 95 °C (2 °C/min). Data were normalized for the housekeeping gene, and the ΔΔCt method was used to determine the fold changes in gene expression with glass coverslips as controls. The primer assays were obtained from QIAGEN as reported in the literature. [24]

**2.5     Statistical analysis**

Statistical significance of the studies was evaluated by the use of GraphPad Prism 6.0 (GraphPad Software, La Jolla, CA, USA) using a one-way ANOVA test followed by Dunnett's multiple comparison test using glass as a control sample. Level of significance was expressed as P-values.

**3.      Results and Discussion**

**3.1     Substrates Characterization**

Optical microscopy and AFM showed the uniformity of thin film coverage on the substrate and no empty areas were observable on the substrates as shown in Fig. **1a-d**. Raman spectroscopy is a useful tool to detect the presence of graphene oxide on any surfaces. The typical spectrum of GO is composed by two peaks: the D peak at ~1350 cm$^{-1}$ and the G peak at ~1586 cm$^{1}$ The intensity ratio ($I_D/I_G$) between these two peaks is employed to characterize GO dispersions. The measured $I_D/I_G$ of GO was 0.92 while upon thermal reduction we noticed a decreased $I_D/I_G$ down to 0.84 (Fig. **1e-f**). XPS C1s spectra of all the different substrates can be deconvoluted into 6 components: C=C (sp$^2$ carbon) at 284.6 eV, C-C (sp$^3$ carbon) at 285.1 eV, C-OH at 286 eV, C-O-C at 286.9 eV, C=O at 287.7 eV and HO-C=O at 288.8 eV (Fig.**1g**) [25,26]. Successful reduction is confirmed by the C1s spectrum of rGO: we noticed a clear decrease in all the oxygen functionalities with the exception of hydroxyl group, a decrease in sp3 carbon and increased sp2 carbon as shown in Fig.**1h**. As previously reported [14], graphene-based materials are efficient materials for the fabrication of artificial scaffolds in regenerative medicine. In fact, graphene and related nanomaterials are characterised by ultra-high specific surface area and by the ability of binding stem cells growth inducers both covalently and non-covalently acting as a pre-concentration platform for growth factors and other biomolecules present in the differentiation medium. Starting form this observation, human dASCs were cultured on the substrates to assess cell proliferation and the effect of the substrate on the gene expression of glial markers.

### 3.2 Proliferation of dASCs on graphene substrates.

To assess dASCs proliferation rate on the different substrates studied we used MTS assay. We selected three time-points on day 1, day 4 and day 7 after seeding. As it can be seen on **Fig. 2a**, the proliferation rate of dASCs on both GO and rGO substrates was comparable to glass coverslips used as controls with values only marginally lower at each time points. To assess the biocompatibility of the substrates and confirm that the slightly reduced proliferation was not due to GO/rGO cytotoxicity we performed a live/dead viability assay after 48 hours of cellular growth on the different substrates. Indeed at each time point, we measured that 99.89 ± 0.01 % of cells were alive on GO substrates, 99.80 ± 0.02 % of cells were alive on rGO coverslips and finally 99.87 ± 0.02 % of cells were alive on glass substrates. We can therefore conclude that although the proliferation rate was marginally slower on the rGO and GO coverslips, the amount of live cells was very high and comparable in all the different substrates studied. We then decided to study the gene expression of crucial proteins and growth factors which are key features of the differentiated state of dASC and are involved in the peripheral nerve regeneration process.

### 3.3  Gene Expression Studies

Key molecules involved in the regeneration process are neurotrophins. These growth factors are involved in neuronal survival, development and functionality [27,28,29]. We decided to focus our attention studying brain-derived neurotrophic factor (BDNF), nerve growth factor

(NGF) and glial-derived neurotrophic factor (GDNF). Also another group of protein molecules involved in the regeneration process are intermediate filament proteins such as nestin and vimentin, which are strictly related to the profound morphological changes associated with SC response to injury.

Nestin is a protein, which is involved in the axonal growth and is normally up regulated after nerve injury [30,31,32,33]. Moreover a recent study highlighted that nestin-positive hair-follicle pluripotent stem cells were able to promote peripheral nerve regeneration [34]. Interestingly, a bigger proportion of ASC were found to express nestin filaments compared to bone marrow mesenchymal stem cells [4]. Vimentin is reported to be up regulated during peripheral nerve regeneration and higher expression levels of this protein have been reported to augment the peripheral nerve regeneration [35]. Lastly we decided to investigate also the gene expression of neurotrophins receptor such as TrkB, TrkC and Ret as a measure of growth factor responsiveness of dASCs.

GDNF expression is increased on rGO coverslips ($1.64 \pm 0.06$ vs glass, $p< 0.01$, n=3) and no statistical difference is observed between GO coverslips and the glass controls. The expression of BDNF is increased on rGO and GO coverslips ($1.53 \pm 0.08$, $p < 0.01$ vs glass, n=3 on rGO substrates) and ($1.68 \pm 0.01$, $p<0.001$ vs glass, n=3 on GO substrates). NGF expression increased on rGO coverslips ($1.67 \pm 0.14$, p-value $< 0.01$) while GO substrates behaved as glass controls. (**Fig. 3 a-c**)

Nestin expression is increased on both rGO and GO substrates ($1.44 \pm 0.05$, $p<0.01$ vs glass, n=3 on rGO substrates) and ($1.31 \pm 0.11$, $p<0.05$ vs glass, n=3 on GO substrates). Vimentin expression is increased on both rGO and GO substrates ($1.66 \pm 0.04$, $p< 0.001$ vs glass, n=3 on rGO substrates) and ($1.67 \pm 0.01$, $p<0.001$ vs glass, n=3 on GO substrates. (**Fig. 3 d-e**)

TrkC expression is increased on rGO substrates ($2.86 \pm 0.29$, $p< 0.01$ vs glass, n=3) while no statistical difference is observed between GO substrates and the glass controls. The same trend is followed by the gene expression of TrkB receptor. The expression of this gene is increased on rGO substrates ($1.85 \pm 0.02$, $p<0.01$ vs glass, n=3). The expression of Ret receptor is marginally increased on rGO substrates although not statistically significant due to the high variability of the rGO substrates. The expression of this gene on GO shows the same behaviour as the glass controls. (**Fig. 3f-h**).

Faroni *et al.* [24] studied the gene expression changes when ASCs are differentiated towards Schwann-like dASCs. The gene expression of GDNF, BDNF, TrkC, Ret, nestin and vimentin markers was reported to be upregulated after the differentiation protocol compared to non-differentiated ASC. The expression of NGF and TrkB was reported to be downregulated after the differentiation protocol compared to non-differentiated ASC, although the level of NGF protein was found to be increased in the dASC phenotype.

The development of a protocol that permanently differentiates ASC cells into dASC is crucial to implement a stem-cell based therapy strategy for peripheral nerve regeneration. dASC cells were found able to express glial markers and to promote nerve regeneration, myelination and to increase the speed of the conduction in the nerve [8-11]. The main obstacle in the clinical translation of dASC is the maintenance of the differentiated phenotype. Faroni et al [24] proved that after the withdrawal of the growth factors, dASC started to decrease the expression of glial markers and to reverse into ASC phenotype. There is consequently the need to develop better differentiation protocols and to test new materials that help maintaining the dASC phenotype even after the withdrawal of the growth factors for efficient delivery of stem cell therapies *in vivo*. Graphene and related nanomaterials have been widely reported as suitable material to support stem cell growth and differentiation [13-17]. In this study, the proliferation rate and the biocompatibility of these substrates were studied by two different assays and we can conclude although there is a slower proliferation rate of dASC on rGO and GO substrates, the amount of live cells is comparable between all the different substrates studies indicating that GO and rGO substrates do not cause cytotoxicity after 48 hours. Importantly, the analysis of gene expression of important glial markers increased after 48 hours of cellular growth on GO and rGO substrates. The expression of neurotrophins and their receptors is statistically increased on rGO substrates and to a lesser extent on GO substrates. Moreover, the expression of intermediate filament proteins such as nestin and vimentin is statistically increased on both and rGO substrates. Park et al. reported increased the neuronal differentiation of neural stem cells on graphene substrates together with decreased expression of glial cells [36]. The contrast between our observations and previous results from Park et al. can be explained by the presence of laminin on the surface of the graphene coverslips. Laminin is a protein of the extracellular matrix which is reported to increase neuronal differentiation of embryonic and neural stem cells [37,38]. Previous results on ASC differentiation on GO substrates [13] showed enhanced osteogenesis, adipogenesis, and epithelial genesis but no experiment was conducted on rGO or graphene substrates. Our study point out the increased glial differentiation especially on rGO substrates. This is very interesting as specific properties of rGO can be exploitable in the clinical translation of dASC. This is especially important for electrical conductivity as rGO compared to GO is not insulating and allows the possibility of electrically stimulate stem cells even in the presence of neuronal co-culture for peripheral nerve regeneration therapeutic strategy.

## 4. Conclusions

Our results confirm the biocompatibility of the graphene-based substrates and show increased expression of neurotrophins and filament proteins mainly on rGO and GO substrates. These results strongly positions rGO and GO coatings to be used as functional surfaces to increase glial differentiation of ASC at earlier stage. As the initial results on 48 hours are encouraging, further studies need to be conducted to establish if the gene expression of these markers will be increased at longer time points as the entire differentiation protocol required 2 weeks of treatment.

**Funding**

AV and AFV acknowledge funding from the Engineering and Physical Sciences research Council (EPSRC) grants EP/G03737X/1 and EP/K016946/1. CS acknowledges funding from Brazilian agency FAPESP (grant 2014/05048-4). AF and AJR are supported by the Hargreaves and Ball Trust, the National Institute for Health Research (II-LA-0313-20003), the Academy of Medical Sciences and the Manchester Regenerative Medicine Network (MaRMN).

**Acknowledgments**

Thank you to Mr. Jonathan Duncan and Miss Siobhan O'Ceallaigh (Consultant Plastic Surgeons), and their patients at the University Hospital of South Manchester for donation of adipose tissue, and to Acorda Therapeutics Inc, for kindly providing GGF-2.

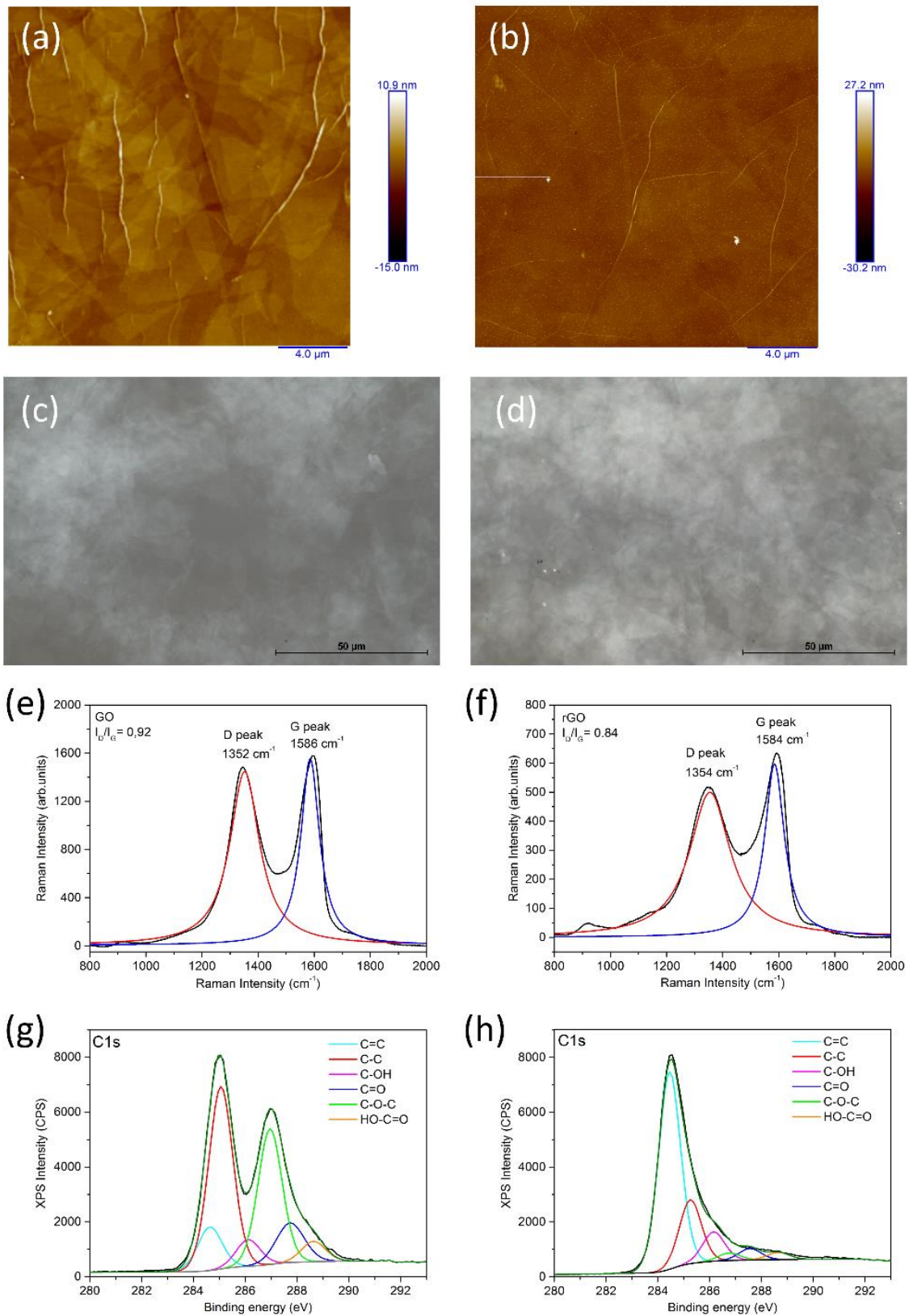

Figure 1: (a, b) AFM topography image of GO, rGO substrates respectively; (c, d) Optical images of GO and rGO substrates respectively; (e, f) Raman spectra of GO and rGO substrates respectively. (g, h) XPS C1s spectra of GO and rGO coated substrates respectively

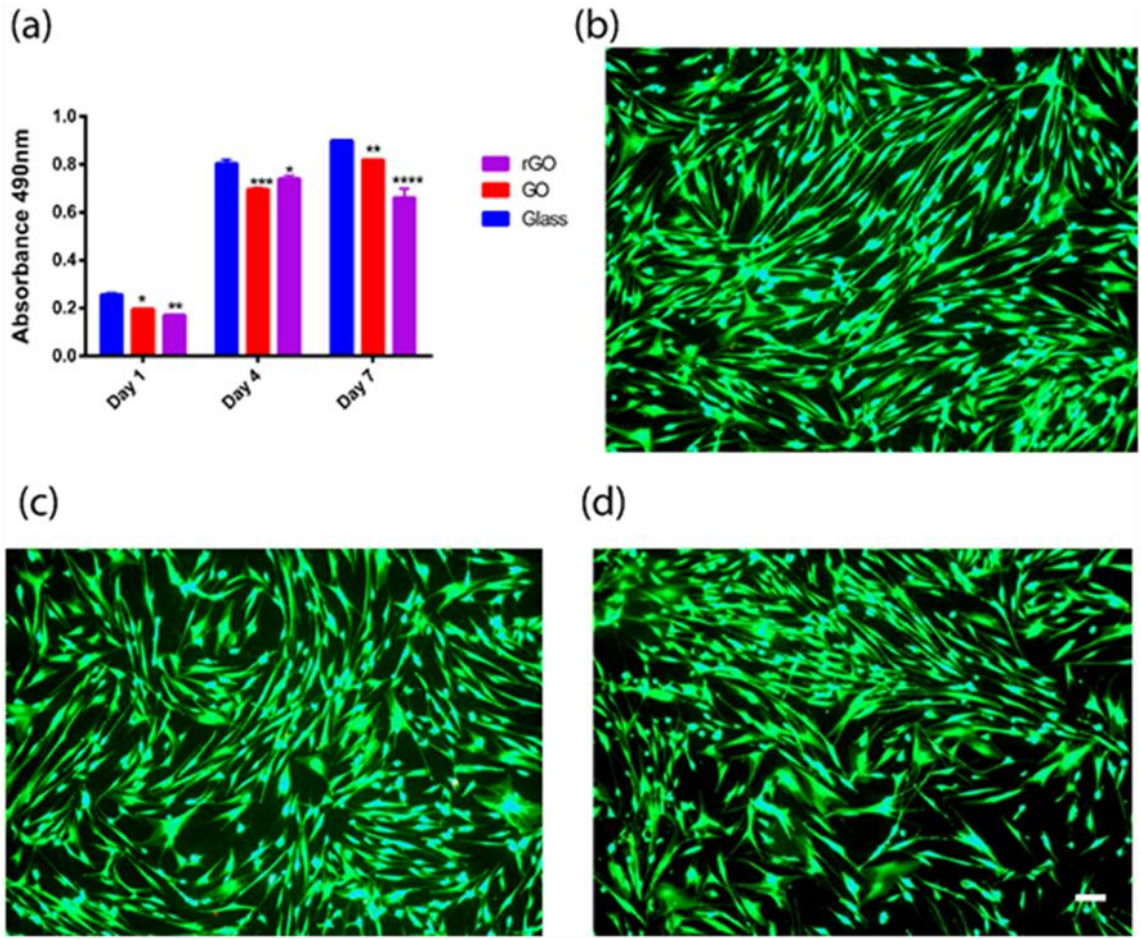

Fig.2: (a) MTS proliferation assay of dASC cells grown on rGO (purple), GO (red) and glass (blue) substrates at day 1, at day 4 and day 7 time-points; **** p<0.0001 ,*** p<0.001, ** p<0.01 , * P<0.05; (b),(c) and (d) dASC cells after 48 hours of cellular growth on glass, GO and rGO respectively. Calcein AM positive cells are alive cells and they are visible in the image due to green fluorescence. 99.89 ± 0.01 % of cells on GO substrates, 99.80 ± 0.02 % of cells rGO coverslips and 99.87 ± 0.02 % of cells on glass substrates were alive. Scale bar: 100 μm

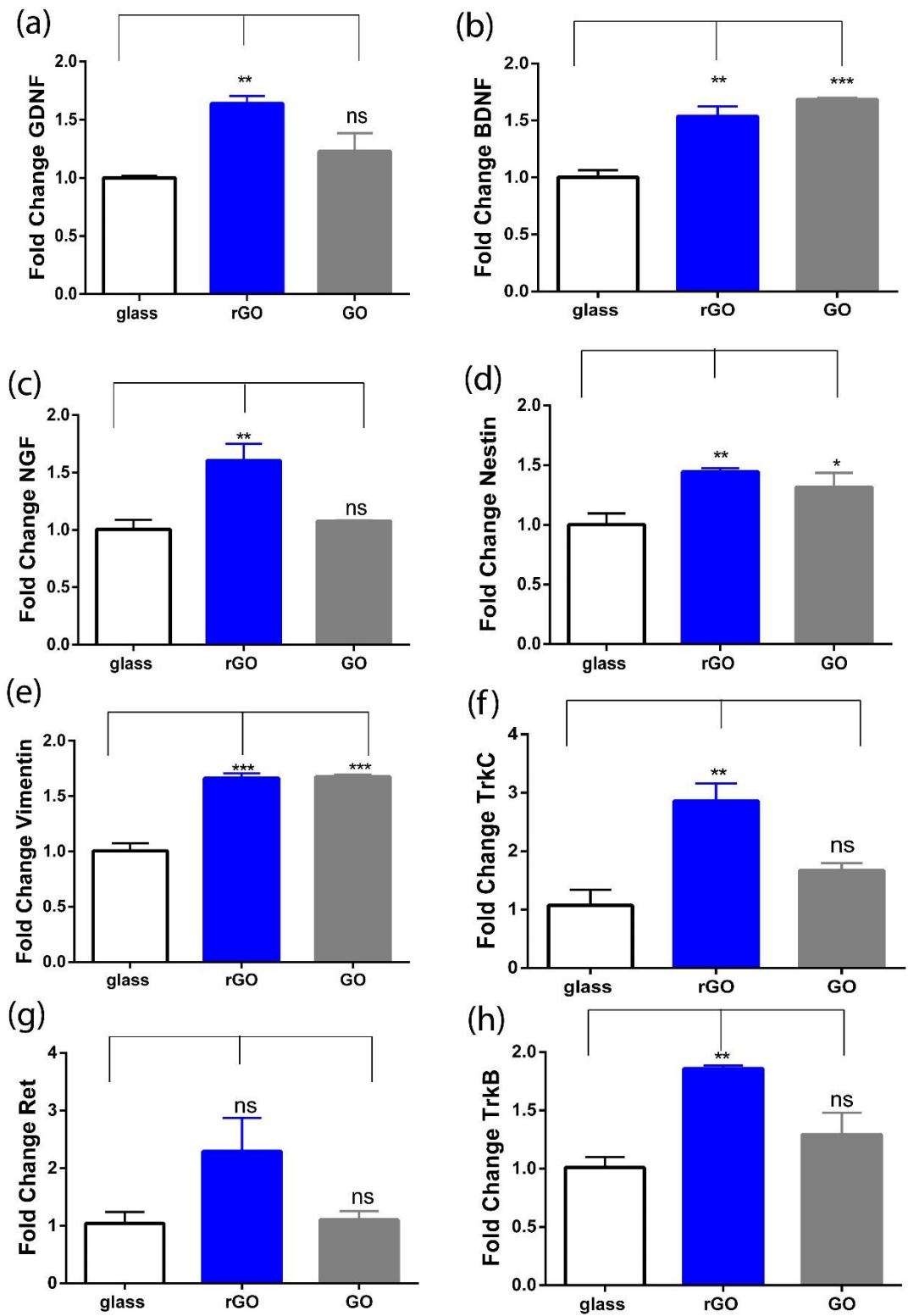

Figure 3: Gene expression after 48 hours of dASCs cellular growth on glass, GO and rGO substrates for the following glial markers:.(a) increased expression of GDNF on rGO substrates; b) increased expression of BDNF on both GO and rGO substrates ; c) increased expression of NGF on rGO; d) increased expression of nestin on both GO and rGO substrates; e) increased expression of vimentin on both GO and rGO substrates; f) increased expression of TrkC receptor on rGO substrates ; g) increased expression of Ret receptor on rGO substrates; h) increased expression of TrkB receptor on rGO substrates. ,*** $p<0.001$, ** $p<0.01$ , * $p<0.05$ and ns= non-significant. Experriments were performed in triplicate.